\pdfoutput=1
\documentclass[acmsmall,screen]{acmart}
\AtBeginDocument{%
  \providecommand\BibTeX{{%
    Bib\TeX}}}

\setcopyright{acmlicensed}
\copyrightyear{2018}
\acmYear{2018}
\acmDOI{XXXXXXX.XXXXXXX}
\acmConference[Conference acronym 'XX]{Make sure to enter the correct
  conference title from your rights confirmation email}{June 03--05,
  2018}{Woodstock, NY}
\acmISBN{978-1-4503-XXXX-X/2018/06}

\usepackage{xspace}
\usepackage{tikz}
\usetikzlibrary{arrows.meta, shapes}
\usetikzlibrary {shadows}
\usetikzlibrary{fit, backgrounds, calc}
\usetikzlibrary{fadings}
\usetikzlibrary{decorations.pathmorphing}

\usepackage{colortbl}
\usepackage{cite}
\usepackage{algorithm}
\usepackage{algorithmicx}
\usepackage[noend]{algpseudocode}
\algrenewcommand\algorithmiccomment[2][\itshape]{{#1\hfill\(\triangleright\)
    #2}}
\algrenewcommand{\algorithmicrequire}{\textbf{Input:}}
\algrenewcommand{\algorithmicensure}{\textbf{Output:}}
\usepackage{graphicx}
\usepackage{textcomp}
\usepackage{fancyvrb}
\usepackage{listings}
\usepackage{xcolor}
\usepackage{float}
\usepackage{todonotes}
\usepackage{booktabs}
\usepackage{multirow}
\usepackage{tcolorbox}
\usepackage{enumitem}
\usepackage{hyperref}
\usepackage{subcaption}

\renewenvironment{quote}
  {\list{}{\leftmargin=0.5cm \rightmargin=0.5cm}%
   \item\relax}
  {\endlist}

\def\BibTeX{{\rm B\kern-.05em{\sc i\kern-.025em b}\kern-.08em
    T\kern-.1667em\lower.7ex\hbox{E}\kern-.125emX}}
    
\usepackage{etoolbox}
\makeatletter
\patchcmd{\@begintheorem}{\textit}{\textbf}{}{}
\makeatother

\usepackage{fontawesome5}
\usepackage{adjustbox}
    
\usepackage{cleveref}
\usepackage{bold-extra} 

\renewcommand{\cite}[1]{\citep{#1}}

\newcommand{\inlineheadingbf}[1]{\medskip\noindent{\bfseries #1.}}
\newcommand{\inlineheadingit}[1]{\medskip\noindent{\it #1.}}

\newcommand{\llbox}[1]{
	\begin{tcolorbox}[width=\columnwidth, colframe=black, boxrule=0.25mm, top=1mm, left=1mm, right=1mm, bottom=1mm]
		#1
	\end{tcolorbox}
}

\DeclareFixedFont{\ttb}{T1}{txtt}{bx}{n}{12} 
\DeclareFixedFont{\ttm}{T1}{txtt}{m}{n}{12}  

\usepackage{color}
\definecolor{deepblue}{rgb}{0,0,0.5}
\definecolor{lightskyblue}{rgb}{0.53, 0.81, 0.98}
\definecolor{deepred}{rgb}{0.6,0,0}
\definecolor{deepgreen}{rgb}{0,0.5,0}

\usepackage{listings}

\definecolor{keyword}{RGB}{0,0,180}
\definecolor{string}{RGB}{163,21,21}
\definecolor{comment}{RGB}{0,128,0}
\definecolor{identifier}{RGB}{0, 0, 0}
\definecolor{number}{RGB}{0, 0, 0}

\newcommand\pythonstyle{\lstset{
    language=Python,
    basicstyle=\ttfamily\footnotesize,
    keywordstyle=\color{keyword}\bfseries,
    stringstyle=\color{string},
    commentstyle=\color{comment}\itshape,
    identifierstyle=\color{identifier},
    numberstyle=\tiny\color{number},
    numbers=left,
    numbersep=8pt,
    showstringspaces=false,
    breaklines=true,
    tabsize=4,
    captionpos=b,
    keepspaces=true,
    morekeywords={self, assert, assume},
}}

\lstnewenvironment{python}[1][]
{
\pythonstyle
\lstset{#1}
}
{}

\newcommand\pythoninline[1]{{\pythonstyle\lstinline!#1!}}

\newcommand{\Pre}{\mathit{Pre}}
\newcommand{\Post}{\mathit{Post}}

\newcommand{\NLC}{\textsc{NL2Contract}\xspace}
\newcommand{\NLPost}{{\em nl2postcond}\xspace}

\newcommand{\pstate}{\sigma}
\newcommand{\PStates}{\Sigma}
\newcommand{\LStates}{\PStates_{gt}}

\begin{document}

\title{Beyond Postconditions: \\Can Large Language Models infer Formal Contracts for Automatic Software Verification?}

\author{Cedric Richter}
\email{cedric.richter@uol.de}
\orcid{0000-0003-2906-6508}
\author{Heike Wehrheim}
\email{heike.wehrheim@uol.de}
\orcid{0000-0002-2385-7512}
\affiliation{%
  \institution{Carl von Ossietzky Universität Oldenburg}
  \city{Oldenburg}
  \country{Germany}
}

\renewcommand{\shortauthors}{Richter and Wehrheim}

\begin{abstract} 
Automatic software verifiers have become increasingly effective at the
task of checking software against (formal) specifications. Yet, their
adoption in practice has been hampered by the lack of such specifications
in real world code. Large Language Models (LLMs) have shown promise
in inferring formal {\em postconditions} from natural language hints embedded in code
such as function names, comments or documentation. Using the generated postconditions 
as specifications in a subsequent verification, however, often leads verifiers 
to suggest {\em invalid}  inputs, 
hinting at potential issues that ultimately turn out to be {\em false alarms}.

To address this, we revisit the problem of specification inference from natural language in
the context of automatic software verification. In the process, we introduce \NLC, the task of employing LLMs 
to translate informal natural language into formal {\em functional contracts}, consisting of postconditions as well as {\em preconditions}. 
We introduce metrics to validate and compare different \NLC approaches, using soundness, bug discriminative 
power of the generated contracts and their {\em usability} in the context of automatic software verification as key metrics. We evaluate 
\NLC with different LLMs and compare it to the task of postcondition generation \NLPost. Our evaluation shows that (1) LLMs are 
generally effective at generating functional contracts {\em sound} for all possible inputs, 
(2) the generated contracts are sufficiently expressive for discriminating buggy from correct behavior, and (3) verifiers supplied
with LLM inferred functional contracts produce fewer false alarms than when provided with postconditions alone. Further investigations show that LLM inferred preconditions
generally align well with developers intentions which allows us to use automatic software verifiers to catch real-world bugs. 
\end{abstract}


\received{11 September 2025}

\maketitle

\section{Introduction}

Software verifiers~\cite{beyer2025improvements, DBLP:conf/popl/BallR02, DBLP:conf/ershov/KhoroshilovMPZ09,  DBLP:conf/cav/BeyerK11, DBLP:conf/tacas/HeizmannCDGHLNM18, DBLP:conf/kbse/PavlinovicLS16} have become increasingly effective at the task of showing that programs adhere to given specifications. 
With the right specification, they can rigorously guarantee the absence of software bugs, verify correctness properties, and significantly reduce the effort and cost associated with manual code inspection and debugging. Yet, the widespread adoption of software verifiers in practice has been largely hindered by the fact that software developers must first of all provide some  formal specification~\cite{snook2001practitioners}. Manually writing these specifications is often highly nontrivial and hence most real world software lacks a {\em formal }specification of what it is intended to do. 

In practice, {\em informal} natural language specifications describing the intended behavior of code, such as code comments or function documentation, are far more common~\cite{pfeiffer2020constitutes}. Software developers use comments and documentation to communicate assumptions over inputs (preconditions) and over the expected behavior of their implementation (postconditions). At the same time, it is well known that software bugs often arise from {\em inconsistencies} between the documented intended behavior and the actual behavior of code~\cite{tan2007icomment, tan2012tcomment, steinbeck2021javadoc}. 
These observations raise the question whether natural language descriptions, which are more common in practice, can be effectively translated into formal specifications, thereby enabling automatic bug detection. As Large Language Models (LLMs) have already proven to be able to synthesize code from natural language intent~\cite{chen2021evaluating, DBLP:conf/iclr/JimenezYWYPPN24}, they are also natural candidates for the technological basis of such a translation. 

Initial investigations~\cite{endres2024can, DBLP:journals/corr/abs-2507-10182} have shown promise in the ability of LLMs to infer formal method postconditions from natural language descriptions of developer intent. Existing works such as \NLPost~\cite{endres2024can} instruct a Large Language Model to write a ``symbolic postcondition'' for a function in the form of an \texttt{assert} statement. The generated postconditions can thus be programmatically checked allowing the automatic validation of their {\em soundness} and {\em completeness}. Still, in practice, we find that the generated postconditions that pass validation are {\em insufficient} when employed in the context of automatic software verification.
More specifically,  when tasked to generate postconditions,  existing LLMs  -- as expected -- typically ignore input assumptions leading to inferred postconditions which  are only valid for a subset of inputs. A consecutively applied verifier then often produces counterexamples, i.e., inputs that violate the postcondition, which ultimately turn out to be {\em false alarms}. 

To address this issue, we revisit in this paper the problem of specification inference from natural language descriptions in the context of automatic software verification. In this process, we propose the task  \NLC as an alternative to \NLPost. 
To this end, we leverage  LLMs for generating {\em functional contracts} in the form of pre- {\em and} postconditions from natural language information within code.
Like for \NLPost, we access  \NLC's ability of generating meaningful contracts using soundness and completeness as our main quality indicators. 
In the process, we especially focus on  \NLC's usage in the setting of software verification,
checking for behavior not matching developer's intents.

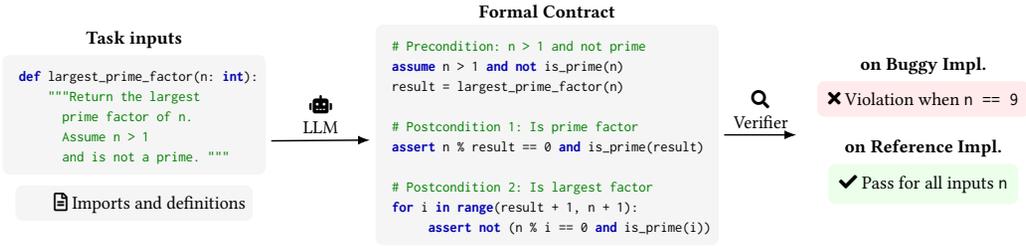
\begin{figure}
\centering
\adjustbox{max width=\textwidth}{
\tikzset{
  warningbox/.style={
    draw=red!80!black,
    fill=red!5,
    very thick,
    rounded corners,
    align=left,
    text width=0.55\linewidth,
    inner sep=6pt,
  },
}

\begin{tikzpicture}

\node[anchor=north west] (code) [fill=gray!8!white, rounded corners, inner sep=4pt] at (0,0) {%
\begin{python}[numbers=none]
def largest_prime_factor(n: int):
    """Return the largest 
      prime factor of n. 
      Assume n > 1 
      and is not a prime. """
\end{python}};
\node [above=0cm of code] (code-header) {\textbf{Task inputs}};
\node[below=0.2cm of code] (add) [fill=gray!8!white, rounded corners, inner sep=4pt] {%
\hspace{0.52cm} \faFile*[regular] ~ Imports and definitions\hspace{0.52cm}};

\node [above right=0cm and 5.5cm of code-header] (contract-header) {\textbf{Formal Contract}};
\node[below=0cm of contract-header] (contract) [fill=gray!8!white, rounded corners, inner sep=4pt] {%
\begin{python}[numbers=none]
# Precondition: n > 1 and not prime
assume n > 1 and not is_prime(n)
result = largest_prime_factor(n)

# Postcondition 1: Is prime factor
assert n % result == 0 and is_prime(result)
    
# Postcondition 2: Is largest factor
for i in range(result + 1, n + 1):
     assert not (n % i == 0 and is_prime(i))
\end{python}};

\node [below right=0.5cm and 4.5cm of contract-header] (mutant-header) {\textbf{on Buggy Impl.}   };

\node[below=0cm of mutant-header] (mutant1) [fill=red!8!white, rounded corners, inner sep=6pt] {%
\faTimes ~ Violation when \texttt{n == 9}
};

\node [below=1cm of mutant-header] (ref-header) {\textbf{on Reference Impl.}   };
\node[below=0cm of ref-header] (ref) [fill=green!8!white, rounded corners, inner sep=6pt] {%
\faCheck ~ Pass for all inputs \texttt{n}
};

\draw($(code.east) + (0.1, -0.5cm)$) edge[-Latex, thick] node[above, text width=1cm, align=center]{\faRobot \\ LLM}  ($(contract.west) + (-0.1, -0.1)$) ;

\draw($(contract.east) + (0.1, 0.0cm)$) edge[-Latex, thick] node[above, text width=1cm, align=center]{\faSearch \\ Verifier}  ($(contract.east) + (1.5cm, 0.0)$) ;

\end{tikzpicture}
}
\caption{Example task for \NLC. The goal is to translate informal natural language descriptions into functional contracts which can be automatically checked by existing software verifiers. 
}\label{fig:motivation}
\vspace{-1.5em}
\end{figure}

\inlineheadingbf{Example} In \Cref{fig:motivation}, we provide an example of an \NLC task adapted from HumanEval/59, a task from the Python code generation benchmark, HumanEval~\cite{chen2021evaluating}. The programmer intends to implement a function (left) that computes the largest prime factor of $n$. In this, she assumes that $n$ is an integer greater than $1$ and not a prime number. Input validation is uncommon in scripting languages such as Python\footnote{\href{https://peps.python.org/pep-0020/}{see PEP 20 - The Zen of Python}} and hence the programmer expects that the user of the function ensures that the assumptions are fulfilled. The goal of \NLC is now to translate the informal natural language specification into a formal contract that can be automatically checked by a verifier. 
Clearly, this contract should capture both input assumptions (as preconditions) and output constraints (as postconditions).
In our example in \Cref{fig:motivation}, we see that \NLC has inferred preconditions (required clauses, given as \texttt{assume}s) and postconditions (ensures clauses, given as \texttt{assert}s) from the function's signature, documentation and possible imports only. 
\NLC evaluates the quality of such inferred contracts by supplying them to an automatic verifier which should then generate bug triggering inputs on buggy implementations. On correct reference implementations, the verifier should return ``pass''. It in particular  should avoid false alarms on correct code.

\inlineheadingit{Limitations of Postcondition Inference} To further motivate \NLC, we discuss the limitations of previous work that target postcondition inference. We use \NLPost as an example. \NLPost evaluates whether LLMs can generate sound postconditions that always hold after the execution of a function. Although postconditions are always dependent on the input and hence could encode precondition constraints, we find that current LLMs are {\em biased} towards simple (incomplete) postconditions that only encode output constraints. For our example in \Cref{fig:motivation} for HumanEval/59, 
 \NLPost generates the following postcondition (taken from \NLPost evaluation): 
\begin{center}
\vspace{-0.25em}
\begin{python}[numbers=none]
        assert n % return_value == 0 \
	        and all(return_value % i != 0 for i in range(2, int(return_value**0.5) + 1))
\end{python}
\vspace{-0.25em}
\end{center}
The postcondition correctly encodes the expectation that the output of the \pythoninline{largest_prime_factor} should be a prime factor of $n$. However, it misses the precondition that $n > 1$ and not a prime number, and misses the postcondition that the result should be the largest prime factor. While the postcondition is test-set correct, i.e., all outputs of the function on test inputs of the evaluation set pass it, we find that software verifiers such as CrossHair~\cite{crosshair} still report false alarms due to the missing input constraints. For example, CrossHair reports a bug for $n = 1$ when evaluated with the given postcondition on the non-buggy reference implementation. Our evaluation will show that this is a shortcoming of the task design of \NLPost when employed within automatic verification, motivating the introduction of \NLC.

\inlineheadingbf{Contributions} 
In this paper, we aim to systematically answer the following overarching research question:
\begin{quote}
{\em Can LLMs translate informal constraints into function contracts useful for software verification?}
\end{quote}
Our interest in (and focus on within evaluation) is the inference of non-trivial and meaningful {\em preconditions}. 
Overall, we provide the following novel contributions:  
\begin{itemize}[leftmargin=0.4cm]
\item We  propose the task of \NLC for {\em inferring functional contracts} (consisting of pre- and postconditions) from developer intent via LLMs. 
\item We develop an {\em exception-based encoding} of functional contracts enabling the usage of automatic verifiers for checking. 
\item We provide sound formal definitions of {\em quality indicators} of functional contracts. 
\item We perform a thorough {\em experimental evaluation} of \NLC itself, including
\begin{itemize}
	\item experiments with different LLMs (GPT-5, GPT-4o, CodeQwen 2.5) and  
	\item a comparison to \NLPost, using both their and our own quality metrics.  
\end{itemize} 
\item  We empirically evaluate the {\em practicality} of generated contracts in the context of automatic software verification. 
\item We provide an evaluation of the usability of inferred specifications for finding {\em real bugs} in the Python-by-Contract benchmark set. 
\end{itemize}

\section{Background}\label{sec:background}
Next, we introduce the background necessary for defining the \NLC task.

\inlineheadingbf{Programs} Throughout this work, we consider simple programs containing functions and statements. We assume that the semantics of a program $P \in \mathbb{P}$ can be modeled as a function $f_{P}: \PStates \rightarrow \PStates$ that maps a program state $\pstate \in \PStates$, e.g. variable assignments, input values etc., to another state $\pstate' \in \PStates$. For each program $P$, we assume that there exists an {\em unknown} ground truth function $f_{gt}: \PStates \rightarrow \PStates$ and set of all {\em valid} input states $\LStates \subseteq \PStates$. In our evaluation, we work with a benchmark set containing known $(f_{gt},\LStates)$ pairs. 
The ground truth function $f_{gt}$ models the {\em correct} intended behavior of $P$ on its domain $\LStates$. A program $P$ is then only {\em correct} iff: 
\begin{equation*}
\forall \sigma \in \LStates: f_P(\sigma) = f_{gt}(\sigma)
\end{equation*}
The program $P$ is {\em buggy} iff it deviates from the intended behavior for at least one valid input state $\sigma_{\text{\faBug}} \in \LStates$:
\begin{equation*}
\exists \sigma_{\text{\faBug}} \in \LStates: f_P(\sigma_{\text{\faBug}}) \neq  f_{gt}(\sigma_{\text{\faBug}}).
\end{equation*} 

\inlineheadingbf{Specifications} Program specifications are a way to define the correct intended behavior of programs, either in its entirety or in parts. In this work, we focus on {\em program contract} based specifications~\cite{meyer1992applying, plosch1997design}. A program contract defines {\em pre-} and {\em postconditions} that the program has to adhere to. Formally, we define a program contract as a pair $(\Pre, \Post)$. A precondition $\Pre: \Sigma \rightarrow \mathbb{B}$ is a predicate specifying for which input states the contract holds. A postcondition $\Post: \Sigma \times \Sigma \rightarrow \mathbb{B}$  specifies the expected output behavior of the given program dependent on the given input.  A  program contract based specification $Spec$ consists of a set of program contracts, i.e. $Spec \subseteq \texttt{PRE} \times \texttt{POST}$  
where $\texttt{PRE}$ is the set of preconditions and $\texttt{POST}$ is the set of postconditions. We let $\mathbb{S}$ be the set of all specifications. 

\inlineheadingit{Soundness} We define a program contract $(\Pre, \Post)$ to be {\em sound} wrt. an ($f_{gt}$, $\LStates$) pair iff: 
\begin{equation*}
 (S1)\quad \forall \sigma \in \Sigma: \Pre(\sigma) \Rightarrow \sigma \in \LStates \hspace{4em} (S2)\quad \forall \sigma \in \LStates: \Pre(\sigma) \Rightarrow \Post(\sigma, f_{gt}(\sigma)) 
\end{equation*}
In other words, a program contract is sound iff every input state $\sigma$ satisfying the precondition is  (1) a valid input state and (2)  leads to an output state (when applying the ground truth function)  jointly satisfying the postcondition.
A specification $Spec$ is {\em sound} wrt. $f_{gt}$ iff it consists of sound program contracts only.  

A sound contract can be used to detect bugs in programs via checking for specification violations. 
A {\em specification violation} of a program $P$ wrt. a specification $Spec$ is an input state $\sigma_{\text{\faTimes}} \in \Sigma$ for which there exists a program contract $(\Pre, \Post) \in Spec$ such that: 
\begin{equation*}
\Pre(\sigma_{\text{\faTimes}}) \land \neg \Post(\sigma_{\text{\faTimes}}, f_{P}(\sigma_{\text{\faTimes}}))
\end{equation*} 
If there exists no such input state, the program $P$ is said to be {\em correct} wrt. $Spec$. 

A specification violation of a {\em sound} specification implies the existence of a bug in the program $P$. More specifically, soundness ensures that the following holds for all $(\Pre, \Post) \in Spec$:  
\begin{equation*}
\forall P \in \mathbb{P}, \sigma \in \PStates:  \quad \Pre(\sigma) \land \neg \Post(\sigma, f_{P}(\sigma)) \quad\Rightarrow\quad \sigma \in \LStates \land f_P(\sigma) \neq f_{gt}(\sigma), 
\end{equation*} 
In other words, a specification violation $\sigma_{\text{\faTimes}}$ always implies the existence of a bug in program $P$ with $\sigma_{\text{\faTimes}}$ as a {\em valid} bug-triggering input.

\inlineheadingit{Completeness} Soundness of a contract thus ensures that specification violations actually hint to bugs. A different quality criterion for contracts is their completeness. Completeness aims at measuring the discriminative power of contracts, i.e., their ability of distinguishing buggy from non-buggy programs via specification violations.  
Soundness  {\em does not} ensure that every bug leads to a specification violation. For this, we additionally need the specification to be {\em sufficiently complete}. Formally, a specification $Spec$ is sufficiently complete wrt. ($f_{gt}, \LStates)$ iff:
\begin{equation*}
(C) \ \ \ \forall P \in \mathbb{P}: \ \  \exists \sigma \in \PStates: \sigma \in \LStates \land f_P(\sigma) \neq f_{gt}(\sigma) \quad\Rightarrow\quad \exists  \sigma_{\text{\faTimes}} \in \PStates: \Pre(\sigma_{\text{\faTimes}}) \land \neg \Post(\sigma_{\text{\faTimes}}, f_{P}(\sigma_{\text{\faTimes}})),
\end{equation*} 
for at least one $(\Pre, \Post) \in Spec$.
Completeness thus ensures that whenever a program $P$ is buggy there exists at least one specification violation.  
A {\em good} specification is both sound and sufficiently complete. In the sequel, we employ soundness and sufficient completeness (and variants thereof) as {\em quality indicators} of specifications. 

\inlineheadingbf{Software Verifier} To find specification violations or prove their absence, we use software verifiers~\cite{beyer2025improvements}. A software verifier checks the correctness of a program $P$ wrt. a specification $Spec$. 
Formally, we view a software verifier as a function ${\sf V}:  \mathbb{P} \times \mathbb{S} \mapsto \{\text{\faCheck}\} \cup \Sigma$
which either returns  \text{\faCheck} (program correct) or an input state for which the program cannot establish the postcondition under the given precondition. Formally, a (sound) verifier should guarantee 
\begin{eqnarray*}
{\sf V}(P, Spec) = \text{\faCheck} & \quad \Rightarrow \quad & P \text{ is correct wrt. } Spec \\ 
{\sf V}(P, Spec) = \sigma & \quad\Rightarrow\quad & \Pre(\sigma) \land \neg \Post(\sigma, f_{P}(\sigma))
\end{eqnarray*}
We first of all employ verifiers for evaluating inferred specifications with respect to soundness and completeness, and ultimately for finding bugs in programs.

\section{\NLC}
\begin{figure}
\centering
\adjustbox{max width=0.95\textwidth}{
\tikzset{
  warningbox/.style={
    draw=red!80!black,
    fill=red!5,
    very thick,
    rounded corners,
    align=left,
    text width=0.55\linewidth,
    inner sep=6pt,
  },
}

\begin{tikzpicture}

\node[anchor=north west] (code) [draw=gray!40!white, fill=gray!8!white, rounded corners, inner sep=0pt, text width=4.2cm] at (0,0) {%
\begin{python}[numbers=none, escapechar=!]
def function_name(!input!):
    """ NL docstring """
\end{python}
};

\node[below=0.4cm of code] (specgen) [fill=violet!12!white, rounded corners, inner sep=4pt, align=center, text width=4cm]  {%
${\sf LLM}_{spec}$
};

\node[below right=0.4cm and -3.7cm of specgen] (spec3) [draw=gray!40!white, fill=gray!8!white, rounded corners, inner sep=0pt, text width=3.7cm]  {%
\vspace{-0.4em}
\begin{python}[numbers=none, escapechar=!]
def contract_3(!input!):
    	...
\end{python}
};

\node[below left=-0.5cm and -3.5cm of spec3] (spec2) [draw=gray!40!white, fill=gray!8!white, rounded corners, inner sep=0pt, text width=3.7cm]  {%
\vspace{-0.4em}
\begin{python}[numbers=none, escapechar=!]
def contract_2(!input!):
    	...
\end{python}
};

\node[below left=-0.5cm and -3.5cm of spec2] (spec1) [draw=gray!40!white, fill=gray!8!white, rounded corners, inner sep=0pt, text width=3.7cm]  {%
\vspace{-0.4em}
\begin{python}[numbers=none, escapechar=!]
def contract_1(!input!):
    	...
\end{python}
};

\node [below right=0cm and -2.7cm of spec1] (sample-header) {\texttt{LLM samples}};
\draw[-Latex, thick] (code) edge node[right]{$\varphi_{nl}$} (specgen);
\draw[-Latex, thick] (specgen) edge node[right]{\small $Spec$} ($(spec3.north) + (-0.3cm, 0.05cm)$);


\node[right=1.2cm of code] (codegen) [fill=violet!12!white, rounded corners, inner sep=4pt, align=center, text width=5cm, minimum height=1cm]  {%
\texttt{ChatGPT completions}
};

\node[below=0.4cm of codegen] (dtest) [draw=gray!40!white, fill=gray!8!white, inner sep=4pt, align=center, text width=5.2cm]  {%
\texttt{differential test} \\
\colorbox{gray!20!white}{ $Pre_{gt}(\sigma) \land f_P(\sigma) \neq f_{gt}(\sigma)$ }
};

\begin{pgfonlayer}{background}
        \node[rounded corners, draw=blue!10, drop shadow=gray!20, fill=lightskyblue!10,  fit=(codegen)(dtest)] (mut-bg) {};
\end{pgfonlayer}

\draw[-Latex, thick] (codegen) edge node[right]{$P$} (dtest);
\draw[-Latex, thick] (code) edge node[above]{$\varphi_{nl}$} (codegen);
\node [above=0cm of mut-bg] (codgen-header) {\texttt{Mutants}};


\node[right=1.1cm of codegen] (gt) [draw=gray!40!white, fill=gray!8!white, rounded corners, inner sep=0pt, text width=4.2cm]  {%
\begin{python}[numbers=none, escapechar=!]
def groundtruth(!input!):
    	...    
\end{python}
};

\node[below=0.1cm of gt] (pregt) [draw=gray!40!white, fill=gray!8!white, rounded corners, inner sep=0pt, text width=4.2cm]  {%
\vspace{-0.4em}
\begin{python}[numbers=none, escapechar=!]
assume pre_gt(!input!)
\end{python}
};

\node[below left=0.2cm and -1.3cm of pregt] (in1) [draw=gray!40!white, fill=gray!8!white, inner sep=4pt]  {%
\small \texttt{input}
};

\node[right=0.3cm of in1] (in2) [draw=gray!40!white, fill=gray!8!white, inner sep=4pt]  {%
\small \texttt{input}
};

\node[right=0.3cm of in2] (in3) [draw=gray!40!white, fill=gray!8!white, inner sep=4pt]  {%
\small \texttt{input}
};

\begin{pgfonlayer}{background}
        \node[rounded corners, draw=green!10, drop shadow=gray!20, fill=green!10,  fit=(gt)(in1)(in3)] (bench-bg) {};
\end{pgfonlayer}
\node [above=0cm of bench-bg] (benchmark-header) {\texttt{Benchmark}};

\draw[-Latex, thick] (bench-bg) -- ($(bench-bg.west) + (-0.32cm, 0.0)$) -- ($(dtest.east) + (0.55cm, 0.0)$) -- (dtest.east);


\node[right=1.2cm of spec2] (sound) [draw=gray!30!white, fill=green!8!white, inner sep=4pt, align=center,]  {%
\colorbox{green!20!white}{ ${\sf V}(P_{gt}, Spec) = \text{\faCheck}$}
};
\node [below=0cm of sound] (sound-header) {\texttt{sound}};

\node[right=0.1cm of sound] (complete) [draw=gray!30!white, fill=red!8!white, inner sep=4pt, align=center,]  {%
\colorbox{red!20!white}{ ${\sf V}(P, Spec) = \sigma_\text{\faTimes}$}
};
\node [below=0cm of complete] (complete-header) {\texttt{complete}};

\node[right=0.1cm of complete] (realbug) [draw=gray!30!white, fill=gray!8!white, inner sep=4pt, align=center,]  {%
\colorbox{gray!20!white}{\small $Pre_{gt}(\sigma_\text{\faTimes}) \land f_P(\sigma_\text{\faTimes}) \neq f_{gt}(\sigma_\text{\faTimes})$}
};
\node [below=0cm of realbug] (realbug-header) {\texttt{bug triggering}};

\draw[-Latex, thick] ($(complete.east) + (-0.2cm, 0cm)$) --  ($(realbug.west) + (0.2cm, 0cm)$);

\begin{pgfonlayer}{background}
        \node[rounded corners, draw=orange!10, drop shadow=gray!20, fill=orange!10,  fit=(sound)(complete)(realbug)(sound-header)] (val-bg) {};
\end{pgfonlayer}
\node [below=0cm of val-bg] (val-header) {\texttt{Verification guided validation}};


\draw[-Latex, thick] ($(spec2.east) + (0.2cm, -0.2cm)$) -- node[above]{$Spec$} (val-bg);
\draw[-Latex, thick] ($(dtest.south) + (0cm, 0cm)$) -- node[right]{$P$} ($(dtest.south) + (0cm, -0.4cm)$);
\draw[-Latex, thick] ($(gt.east) + (0.05cm, -0.3cm)$) -- ($(gt.east) + (0.6cm, -0.3cm)$) --   ($(val-bg.east) + (0.4cm, 0cm)$) -- ($(val-bg.east)$);

\end{tikzpicture}
}
\caption{Overview of \NLC's specification validation process.}\label{fig:overview}
\vspace{-1em}
\end{figure}
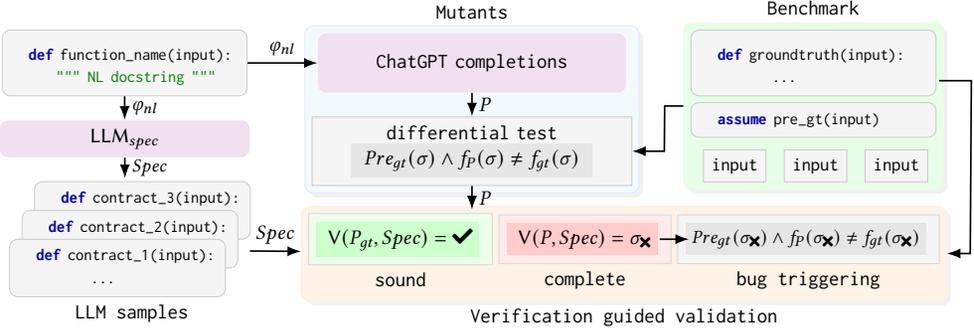

In this section, we introduce the \NLC task and how we leverage it to evaluate the capabilities of LLMs to generate {\em sound}
and {\em complete} specifications. \Cref{fig:overview} provides an overview of our evaluation approach.  

\subsection{Task Formulation}\label{sec:task}
The goal of \NLC is to evaluate the ability of LLMs to generate specifications $Spec$ {\em consistent} with a natural language description $\varphi_{nl}$. Formally, we model the LLM as a function ${\sf LLM}_{spec}: \Phi_{nl} \rightarrow \mathbb{S}$ which maps a natural language description  $\varphi_{nl} \in \Phi_{nl}$ to a formal specifications $Spec \in \mathbb{S}$. The key question is how to measure that $Spec$  is consistent with $\varphi_{nl}$. For this, we propose the following task design. 

\inlineheadingbf{Task Design} An \NLC task is a triple $(\varphi_{nl}, P_{gt}, \Pre_{gt})$ of natural language description $\varphi_{nl}$, reference implementation $P_{gt}$ and reference precondition $\Pre_{gt}$, the latter for instance given as assertions within programs. The program $P_{gt}$ implements the ground truth function $f_{gt}$ and the precondition $\Pre_{gt}$ describes its set of valid inputs $\LStates$.

The goal of ${\sf LLM}_{spec}$ when given an \NLC task is to infer a specification $Spec$ from the natural language description $\varphi_{nl}$.  Naturally, the LLM  neither has access to the reference implementation nor to the precondition. 
We use $P_{gt}$ and $\Pre_{gt}$ to evaluate the quality of the inferred specification $Spec = {\sf LLM}_{spec}(\varphi_{nl})$.
As detailed in \Cref{sec:background}, a high quality contract has to be sound and sufficiently complete wrt.~an ($f_{gt}$, $\LStates$) pair 
which our task contains in the form of $(P_{gt}, \Pre_{gt})$. 

\inlineheadingbf{Verification Guided Validation} Looking at the definitions of soundness and (sufficient) completeness, we see two challenges when actually evaluating contracts: Soundness needs to be checked on {\em all} states and completeness additionally investigates {\em all} programs. In particular the latter is infeasible in practice. Here, we address the first challenge by employing a formal verifier, ideally {\em proving} specification soundness, and a combination of verifier and mutations for the second challenge. 

\inlineheadingit{Soundness} For soundness we evaluate requirements (S1) and (S2) jointly. 
More specifically, let  $(\Pre,\Post) \in Spec$ be one contract of a specification inferred by ${\sf LLM}_{spec}$. 
Instead of separately checking (S1) and (S2), we derive a new specification $Spec_{*} = \{(\Pre, \Post_{*}) \mid (\Pre, \Post) \in Spec\}$ with $\Post_{*}(\sigma, \sigma') = \Pre_{gt}(\sigma) \land \Post(\sigma, \sigma')$ for our evaluation. 
We then ask the verifier ${\sf V}$ to check whether $P_{gt}$ satisfies $Spec_{*}$. If  ${\sf V}(P_{gt}, Spec_{*}) = \text{\faCheck}$, we have 
\begin{equation*}
\forall \sigma \in \Sigma: \Pre(\sigma) \Rightarrow \Pre_{gt}(\sigma) \land \Post(\sigma, f_{gt}(\sigma))
\end{equation*} 
 for all $(\Pre, \Post) \in Spec$ and thus know soundness of $Spec$. By this construction, we can check soundness within a single verifier call.

\inlineheadingit{Completeness} While the verifier will often be able to provide a proof of properties (S1) and (S2) for all states, 
it will -- naturally -- not reason about {\em all} possible programs. Instead of checking property (C) on all programs, we thus employ mutations, more precisely buggy variants of the reference implementation (alike~Endres et al.\cite{endres2024can}), to approximate completeness.   Here, the key idea is to generate a large set of {\em buggy} code mutants $CM$ that show different behavior than the reference implementation, i.e.:
\begin{equation*}
CM(P_{gt}) \subseteq \{ P \in \mathbb{P} \mid \exists \sigma \in \PStates: \Pre_{gt}(\sigma) \land f_{P}(\sigma) \neq f_{P_{gt}}(\sigma)\}
\end{equation*} 
We then evaluate  completeness on this set of code mutants $CM$, i.e., for every $P \in CM$, the verifier is tasked to find specification violations. If  we get ${\sf V}(P,Spec) =  \sigma_{\text{\faTimes}}$ (for some $\sigma_{\text{\faTimes}} \in \Sigma$) for all $P \in CM$, we consider $Spec$ to be sufficiently complete. Recall that in case of sound 
specifications this moreover guarantees $\sigma_{\text{\faTimes}}$ to be a bug triggering input. 
In practice, $CM(P_{gt})$ can be constructed by modifying the reference implementation. In our evaluation, we use code mutants generated by ChatGPT tasked to find code completions for the natural language description $\varphi_{nl}$.

\subsection{Prompt Design for LLM-Based Contract Inference}\label{sec:contract}
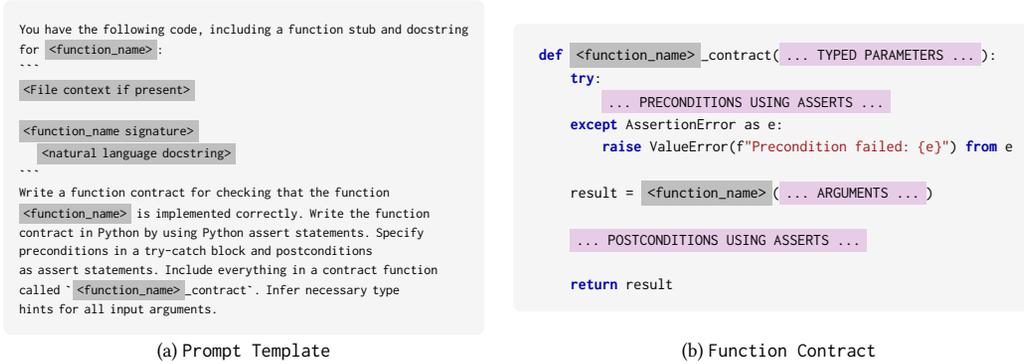
\begin{figure}
\centering
\adjustbox{max width=\textwidth}{
\tikzset{
  warningbox/.style={
    draw=red!80!black,
    fill=red!5,
    very thick,
    rounded corners,
    align=left,
    text width=0.55\linewidth,
    inner sep=6pt,
  },
}

\begin{tikzpicture}

%
%
%

\node[anchor=north west] (code) [fill=gray!8!white, rounded corners, inner sep=8pt, text width=8cm] at (0,0) {%
\adjustbox{max width=1\textwidth}{%
\begin{lstlisting}[escapechar=!, numbers=none, basicstyle=\ttfamily\small]
You have the following code, including a function stub and docstring 
for !\colorbox{gray!50}{<function\_name>}!:
```
!\colorbox{gray!50}{<File context if present>}!

!\colorbox{gray!50}{<function\_name signature>}!
	!\colorbox{gray!50}{<natural language docstring>}!
```
Write a function contract for checking that the function 
!\colorbox{gray!50}{<function\_name>}! is implemented correctly. Write the function 
contract in Python by using Python assert statements. Specify
preconditions in a try-catch block and postconditions 
as assert statements. Include everything in a contract function 
called `!\colorbox{gray!50}{<function\_name>}!_contract`. Infer necessary type 
hints for all input arguments. 
\end{lstlisting}
}};
\node [below=0cm of code] (code-header) {(a) \texttt{Prompt Template}};

\node[right=0.5cm of code] (format) [fill=gray!8!white, rounded corners, inner sep=8pt]  {%
\adjustbox{max width=0.8\textwidth}{%
\begin{python}[escapechar=!, numbers=none]
def !\colorbox{gray!50}{<function\_name>}!_contract(!\colorbox{violet!20!white}{... TYPED PARAMETERS ...}!):
    try:
        !\colorbox{violet!20!white}{... PRECONDITIONS USING ASSERTS ...}!
    except AssertionError as e:
        raise ValueError(f"Precondition failed: {e}") from e
    
    result = !\colorbox{gray!50}{<function\_name>}!(!\colorbox{violet!20!white}{... ARGUMENTS ...}!)

    !\colorbox{violet!20!white}{... POSTCONDITIONS USING ASSERTS ...}!

    return result
\end{python}
}};
\node [right=6cm of code-header] (format-header) {(b) \texttt{Function Contract}};


%
%
%
%
%
%

\end{tikzpicture}
}
\caption{Prompt template for generating functional contracts from natural language descriptions. Dark gray parts are automatically filled in during prompt construction. Violet parts are inferred by the LLM from the given context. }\label{fig:prompt}
\vspace{-1.2em}
\end{figure}

We next discuss our prompt design for instructing LLMs to generate program contracts from natural language descriptions. \Cref{fig:prompt} shows our prompt approach. The LLM is instructed to generate function contracts from developer hints and given context. 
A function contract specifies the behavior of a single function or method. Here, we specifically look at function contracts that specify the behavior of Python functions. 

\inlineheadingbf{Prompt Construction} We construct the prompt for the LLM directly from existing code. For this, we take as input the function signature, a natural language docstring, and the surrounding file context (if available). We then use this information to automatically construct the prompt shown in \Cref{fig:prompt}(a). In practice, such information is often not readily available.  Therefore, when employing \NLC approaches for finding bugs in real world code with automatic software verifiers, we parse this information directly from source code.  Given a function under test, we first identify all functions (methods) called and all variables accessed in the global or class context.  Then, we construct the file context by (1) including variable definitions of variables defined in the global or class context, (2) signatures and documentation of called methods and functions, and (3) the header of the class (if included in a class). We have found that including this information is important for LLMs to generate valid contracts -- otherwise LLMs tend to hallucinate non-existing code.

\inlineheadingbf{Function Contracts in Python} While there are many attempts to bring formal contracts into the Python language such as \texttt{icontract}~\cite{icontract}, \texttt{deal}~\cite{deal} or the abandoned \href{https://peps.python.org/pep-0316/}{PEP 316}, there is still only limited supported from automatic verification tools. 
Note in particular that there is no natural support of \texttt{assume} statements like we used for illustration in the example of \Cref{fig:motivation}. 
For this reason, we decided to encode pre- and postconditions as pure Python assertions which allows us to use software verifiers subsequently checking for assertion violations. Our encoding is shown in \Cref{fig:prompt}(b). The contract is encoded as Python function that takes in the same parameters as the original function (and calls it). 
The encoding should guarantee that an assertion exception (hinting to a specification violation) occurs for inputs which satisfy the precondition but where the function's returned value (in \Cref{fig:prompt}(b): \texttt{result}) does not meet the postcondition. 
To this end, preconditions are encoded as Python \texttt{assert} statement that are wrapped in a try-catch block (\texttt{try-except}).   The purpose of the try-catch block is to prevent precondition violations to lead to assertion errors. The function then instead raises a \texttt{ValueError} and stops. 
The postconditions can directly be encoded as \texttt{assert} statements which are dependent on the input and the result of the original function. The function hence raises an assertion error only for function inputs that satisfy the precondition (and are hence valid) and violate the postcondition which exactly coincides with our definition of specification violation.  

In practice, we provide this function contract template as part of the prompt by filling in the function name of the original function. The LLM then infers the missing preconditions (which also includes the types of function parameters) and postconditions from the given natural language context.

\inlineheadingbf{Function Contracts in other programming languages} Our encoding of functional contracts can be easily adopted for programming languages that support exception handling. In particular in Java,  pre- and postconditions can be encoded as Java assertions (which are available since Java 1.4) and the whole contract as a Java method. \texttt{IllegalArgumentException} can be raised for precondition violations which is a native way to indicate invalid inputs. In general, we see our encoding of functional contracts as a feasible way to evaluate LLM's ability to generate functional contracts across many programming languages. We however focus in our evaluation on Python, and leave the evaluation in other languages open for future work.

\section{Evaluation}
We perform a systematic study to evaluate the ability of LLMs to infer specifications from natural language intent on popular benchmarks. During our evaluation, we thoroughly  evaluate whether LLMs can generate {\em sound} and {\em complete} specifications that enable the automatic detection of software bugs with automatic software verifiers. 
Our evaluation is guided by the following research questions:
\begin{description}
\item[\textbf{RQ1}] How effective are LLMs in generating {\em sound} functional contracts?
\item[\textbf{RQ2}] How effective are LLM-generated specifications in {\em discriminating} buggy and correct behavior?
\item[\textbf{RQ3}] Are LLM-generated specifications useful for finding real bugs {\em automatically}?
\end{description}
In \textbf{RQ3}, we explore the usability of LLM-generated specifications in different verification scenarios, including their potential for automatic testing.

\subsection{Experimental Setup}
In the following, we describe the main experimental setup for \textbf{RQ1} and \textbf{RQ2}. We see \NLPost as our main baseline, but we compare also with other baselines on the task of bug detection in our experiments for \textbf{RQ3}. Our experimental setup for \textbf{RQ3} is provided in \Cref{sec:rq3}. 

\inlineheadingbf{Benchmark} We adopt {\textsc{HumanEval}$^+$}~\cite{liu2023your} as our main benchmark. {\textsc{HumanEval}$^+$} includes 164 Python problems, each consisting of a function stub, a natural language description encoded as a Python docstring, a reference implementation, and some validating tests. We choose {\textsc{HumanEval}$^+$} as it updates the popular {\textsc{HumanEval}} benchmark~\cite{chen2021evaluating} with (1) a more extensive test suite (775 tests per problem on average), (2) corrected reference implementations, and most importantly (3) {\em reference preconditions} implemented as Python asserts for each problem in the benchmark. The preconditions are non-trivial: 92 out of 164 tasks contain preconditions that are not type related (e.g. \pythoninline{assert not is_prime(n)}). A total of 46 out of 164 tasks are annotated with preconditions that use functions from the Python standard library (e.g. \pythoninline{assert set(a).issubset(\{"0", "1"\})}) and  5 out of 164 tasks are annotated with loops, conditional statements, or introduce helper functions to encode preconditions. In addition, 104 out of 164 tasks do not provide a type hint for input parameters in the function stub, which makes them generally challenging for automatic software verification. Based on these observations, we see {\textsc{HumanEval}$^+$} as a demanding benchmark that allows to evaluate whether LLMs can generate functional contracts from natural language descriptions that are also {\em sound} with respect to the reference solution and precondition. 

\inlineheadingbf{Large Language Models} We generate functional contracts with recent large language models that have shown strong performance on various programming tasks. We include both open-source and closed source models: 
\begin{description}[leftmargin=0em]
\item[GPT-5 (Chat)] and \textbf{GPT-4o} are the most recent version GPT series of chat models provided by OpenAI. GPT-5 shows strong performance on existing coding benchmarks~\cite{DBLP:conf/iclr/JimenezYWYPPN24}. 

\item[CodeQwen 2.5] CodeQwen 2.5 (32B) is the currently best 32B parameter open-source coding model\footnote{\href{https://evalplus.github.io/leaderboard}{https://evalplus.github.io/leaderboard}, accessed in September 2025.} that still fits on consumer hardware. We use a variant of CodeQwen 2.5 that is tuned for instruction following 
which supports the chat format for assistance in coding. 
\end{description}
We use the same prompt for all coding models and use the OpenRouter API~\footnote{\href{https://openrouter.ai}{openrouter.ai}} to query all models. We see the performance of these models as a baseline performance which can likely be improved by further prompt tuning or other more advanced techniques~\cite{wei2022chain}.

\inlineheadingbf{Contract Generation} For each {\textsc{HumanEval}$^+$} problem, we generate 10 functional contracts per problem and model. We use the default temperature of $0.7$ as this has been found to be reasonable for code generation tasks. As our baseline prompt, we employ \NLPost~\cite{endres2024can}. As \NLPost instructs the LLM to generate postconditions, we wrap the generated postconditions in a contract function to be evaluated in our setup: 
\begin{center}
\vspace{-0.5em}
\begin{python}[escapechar=!, numbers=none, xleftmargin=.2\textwidth ]
def !\colorbox{gray!50}{<function\_name>}!_contract(!\colorbox{gray!50}{<original function\_signature>}!):
    return_value = !\colorbox{gray!50}{<function\_name>}!(!\colorbox{gray!50}{<arguments>}!)
    !\colorbox{violet!20!white}{... POSTCONDITION GENERATED BY \NLPost~ ...}!
    return return_value
\end{python}
\vspace{-0.5em}
\end{center}
We automatically fill in the function name and signature based on the given programming task. Note that postconditions generated by LLMs with the \NLPost prompt expect the variable \texttt{return\_value} to contain the result of the computation which we compute by calling the original function. We view the function signature as a {\em trivial precondition} that can easily be computed from the function code. The verifier will use the type hints to restrict the search space for finding specification violations. Finally, we generate in total 9,840 functional contracts across all models, prompt variants (\NLC and \NLPost), and programming problems.

\inlineheadingbf{Verification} To verify the soundness and completeness of the generated specifications, we employ automatic verification tools. There are currently only a few formal software verifiers~\cite{crosshair, DBLP:conf/cav/Eilers018, DBLP:conf/issta/0001MFSC24} that support the verification of Python code. In our experiments, we specifically decided for CrossHair~\cite{crosshair} which we run with the analysis option \texttt{----analysis\_kind=asserts}\footnote{CrossHair expects that all functions to be verified are marked with an assert statement in the beginning and it reports all types of exceptions as errors. Hence, when we run our contracts with CrossHair, we preprocess them by adding a trivial assertion (\texttt{assert True}) in the beginning and we replace the raising of  \texttt{ValueError} with a \texttt{return} statement.}. CrossHair is a symbolic execution engine for Python and it supports the verification of programs with basic Python types. We choose CrossHair because it is capable enough for verifying  {\textsc{HumanEval}} tasks and because it is generally effective in finding assertion errors in Python programs. However, CrossHair -- as most automatic verifiers for Python --  is {\em incomplete}, i.e., it reports \texttt{unknown} results if it cannot prove the complete input space to be safe. Therefore, we adopt a more pragmatic approach and declare a specification as unsound if CrossHair finds a specification violation on the reference implementation in 60s and sound otherwise. 

\inlineheadingbf{Code Mutants} To evaluate specification completeness, we employ the same set of code mutants $CM$ as used in the evaluation of  \NLPost postconditions. The set contains 4 to 233 code mutants per {\textsc{HumanEval}} task with a median of 55 mutants per task. The mutants are generated as code completions from an earlier version of ChatGPT. Each code mutant comes with a set of  bug triggering inputs, i.e. {\em valid} inputs for which the mutant behaves differently than the reference implementation. 

\inlineheadingbf{Metrics} Our key metrics for the evaluation of LLM inferred specifications are soundness and completeness, checked via verifiers according to 
our description  in \Cref{sec:task} (thus called {\em verification soundness} and {\em verification completeness}).  
 For a set of $n$ randomly generated specifications, we compute $sound@k$ as a $pass@k$ style metric~\cite{chen2021evaluating} for $1 \leq k< n$ that measures the statistical expected value that a random sample of $k$ specifications contains at least one verification sound specification. This is in line with the $accept@k$ metric used in the \NLPost evaluation to estimate test-set correctness.
 For completeness, we compute $\% complete$ as the expected value that a verification sound specification is also verification complete, i.e., the verifier kills all mutants given the specification. For our comparison with \NLPost, we furthermore employ {\em test-set correctness} and {\em bug completeness} as defined by Endres et al.~\cite{endres2024can}. These metrics are computed by using the set of test inputs associated with each {\textsc{HumanEval}$^+$} task instead of employing a verifier to check {\em all} inputs.

\section{Results}
We now discuss the results of our experimental evaluation per research question.
 
 \subsection{RQ1: Ability of LLMs to Generate Sound Functional Contracts}
 To answer \textbf{RQ1}, we evaluate the effectiveness of LLMs to generate sound functional contracts. We compare models and prompts with respect to test-set correctness and verification soundness.
 
 \begin{table*}
  \caption{Comparison of different LLMs evaluated with the \NLC and \NLPost prompt with respect to test-set correctness and verification soundness, all in \%. The best results per model are highlighted in bold.}
  \label{tab:sound-results}
  \centering
  \begin{adjustbox}{max width=\textwidth}
  \newcolumntype{g}{>{\columncolor{blue!10}}c}
  \newcolumntype{b}{>{\columncolor{blue!10}}c}
  \begin{tabular}{l c bg g gg g gg g gg g gg g gg }
    \toprule
    \rowcolor{white}
    && &&  \multicolumn{5}{c}{\bfseries Test-set correctness} &&  \multicolumn{5}{c}{\bfseries Verification soundness} \\
    \cmidrule{4-9} \cmidrule{11-15} \rowcolor{white}
    Model && Prompt && $accept@1$ && $accept@5$ && $accept@10$ && $sound@1$ && $sound@5$ && $sound@10$ \\
    \midrule
    \rowcolor{white}
    \multirow{ 2}{*}{GPT-5 (Chat)} && \NLPost && 86.6 && 94.0 && \textbf{95.7} && 13.0 && 14.3 && 15.2 \\
    && \NLC && \textbf{89.0} && \textbf{94.6} && 95.1 && \textbf{81.1} && \textbf{88.7} &&  \textbf{90.0} \\
    \midrule
    \rowcolor{white}
    \multirow{ 2}{*}{GPT-4o} && \NLPost && 75.9 && 92.2 && 95.1 && 13.4 && 16.6 && 18.3 \\
    && \NLC &&  \textbf{80.5} &&  \textbf{93.5} &&  \textbf{96.3} && \textbf{72.8} && \textbf{87.9} && \textbf{91.5} \\
    \midrule
    \rowcolor{white}
    \multirow{ 2}{*}{CodeQwen 2.5} && \NLPost && 59.1 && 80.1 && 85.4 && 11.5 && 15.5 && 17.1 \\
    && \NLC &&  \textbf{73.6} && \textbf{94.8} && \textbf{97.0} && \textbf{66.7} && \textbf{88.7} && \textbf{93.9} \\

    \bottomrule
  \end{tabular}
  \end{adjustbox}
  \vspace{-1em}
\end{table*}
 
 \inlineheadingbf{Results} \Cref{tab:sound-results} summarizes our experimental results. Overall, we find that for {\textsc{HumanEval}$^+$}: 

\inlineheadingit{LLMs are effective in generating test-set correct specifications from natural language} We observe that the evaluated LLMs produce test-set correct functional contracts in between 73.6\% to 89.0\% of all cases ($accept@1$). If we sample more specifications, it is highly likely that the generated functional contract is test-set correct (up to 97\% for $accept@10$). When comparing the same LLM with different prompt variations, we find that the performance of the OpenAI models do not suffer from performing the more complex tasks of contract generation. In fact, for $accept@1$, they perform significantly better in generating test-set correct functional contracts with \NLC than generating a test-set correct postcondition with \NLPost. Surprisingly, the open-source model CodeQwen 2.5 performs in general better in generating test-set correct contracts than generating test-set correct postconditions. We find that LLMs prompted to generate functional contracts tend to split postconditions into multiple simpler assertions which are more often test-set correct.

\begin{figure}[t]
    \centering
    \begin{subfigure}{0.5\textwidth}
       \centering
         \includegraphics[width=\textwidth]{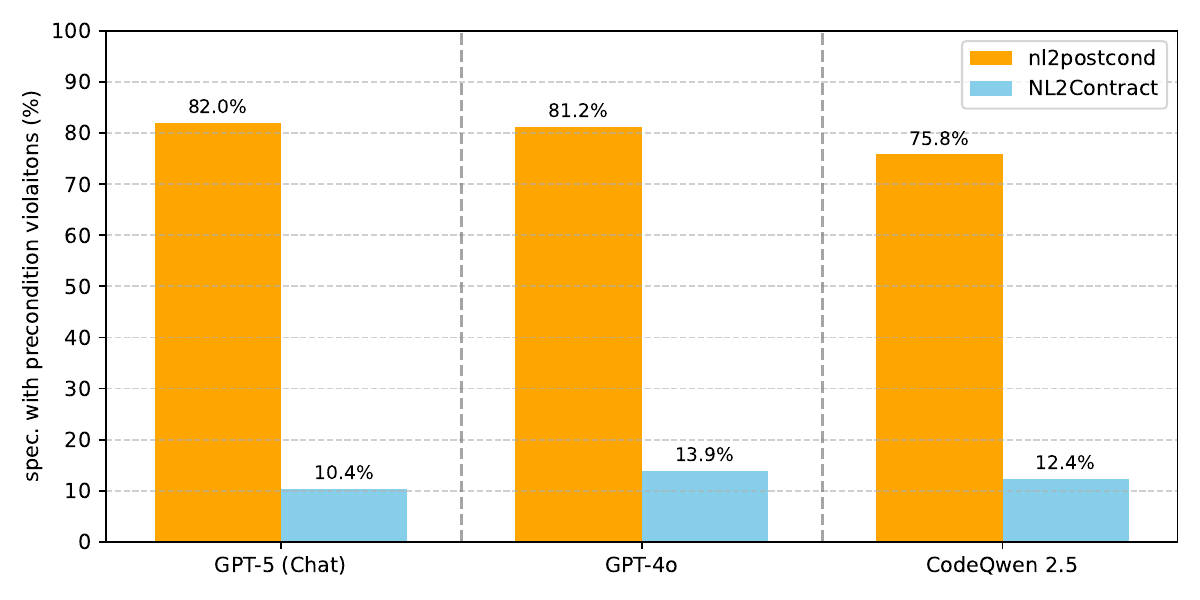}
  	\caption{Precondition unsoundness (lower is better)}
  	\label{fig:pre-sound}
    \end{subfigure}\hfill
    \begin{subfigure}{0.5\textwidth}
        \centering
         \includegraphics[width=\textwidth]{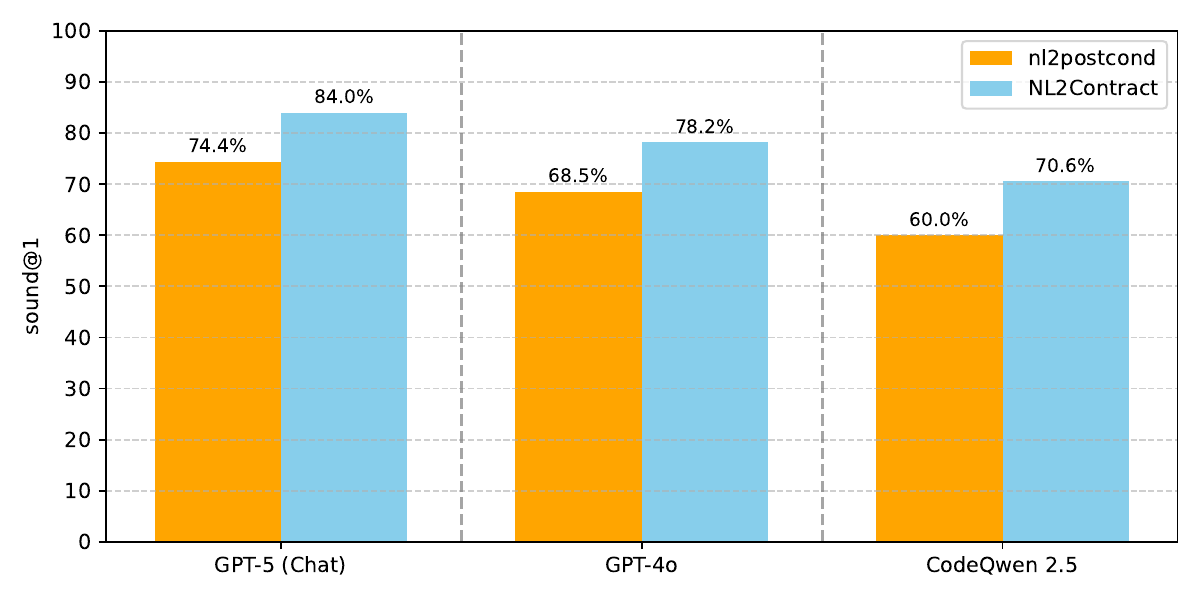}
  	\caption{Postcondition $sound@1$ (higher is better)}
  	\label{fig:post-sound}
    \end{subfigure}
    \vspace{-1.5em}
    \caption{Comparison of LLMs under different prompts. Left (a) shows the percentage of specifications that are unsound due to violations of reference precondition. Right (b) shows $sound@1$ of the postconditions. }
    \label{fig:prepost-sound}
 
  \vspace{-1em}
\end{figure}

\inlineheadingit{Test-set correctness does not imply verification soundness} We observe that the task of generating verification sound specifications (right hand side of \Cref{tab:sound-results}) is significantly more challenging than producing test-set correct specifications. The performance of LLMs prompted with \NLPost drop drastically from 73.6\% -- 89.0\% $accept@1$ test-set correctness to 11.5\% -- 13.0\% $sound@1$ when evaluated with respect to verification soundness. LLMs prompted to generate functional contracts with \NLC achieve a verification soundness score $sound@1$ of 66.7\% to 81.1\%. By increasing the sampling budget, LLMs become more effective in generating verification sound functional contracts ($sound@5$ of 87.9\% to 88.7\% and $sound@10$ of 90.0\% to 93.9\%). To better understand why LLMs are more effective in generating verification sound functional contracts, we analyze the distribution of soundness violations. \Cref{fig:pre-sound} shows the percentage of specifications (out of all generated specifications) that are unsound due to a precondition violation for each model and prompt. We observe that most specifications generated with \NLPost (75.8\% to 82.0\%) are unsound due to violations of condition (S1) of \Cref{sec:background}, i.e. they allow {\em invalid} inputs that violate the reference precondition. LLMs prompted with \NLC are better in capturing the ground truth precondition. Only 10.4\% to 12.4\% of the \NLC specifications are unsound due to violations of the reference precondition. We will see in \Cref{sec:rq3} that this has a significant practical impact: Software verifiers that are supplied with verification unsound specification often produce a significant higher number of false alarms. Only verification soundness indicates this shortcoming of \NLPost postconditions. This shows the importance of quality metrics in software verification that evaluate the soundness of specifications for {\em all} possible inputs. 

\inlineheadingit{Postcondition soundness} We also explore and compare the verification soundness of the generated postconditions. 
To this end, we add the reference precondition as an additional precondition to the generated contracts, both for \NLC and \NLPost. This ensures that the verifier only generates {\em valid} inputs when validating the contracts. Hence, the contracts can only be unsound due to violations of condition (S2), i.e. the postcondition fails on the reference implementation given a valid input. Our $sound@1$ results for this experiment are shown in \Cref{fig:post-sound}. LLMs prompted for \NLC achieve a postcondition soundness score $sound@1$ score of 70.6\% to 84.0\%. This is on average 9,6\% to 10,6\% higher than the $sound@1$ score achieved by LLMs prompted with \NLPost. 

Based on these results, we conclude for \textbf{RQ1}:
\llbox{On {\textsc{HumanEval}$^+$} tasks, LLMs are effective in generating {\em test-set correct} and {\em verification sound} functional contracts from informal natural language specifications. LLMs with \NLC are significantly better in producing verification sound specifications than LLMs prompted for \NLPost. LLMs prompted with \NLC are effective in capturing input assumptions. }

\subsection{RQ2: Ability to Discriminate Buggy and Correct Behavior}
To answer \textbf{RQ2}, we evaluate whether the generated {\em verification sound} specifications can discriminate buggy and correct behavior. We compare models and prompts with respect to (test-set) bug completeness and verification completeness.
 \begin{table*}
  \caption{Comparison of different LLMs evaluated with the \NLC and \NLPost prompt with respect to bug completeness and verification completeness. $\%complete$ is the percentage of verification sound specification that are also complete (kills all mutants). $x/164$ is the number of tasks (out of 164) with at least one complete specification. $\%killed$ is the percentage of mutants killed by the verifier with at least one verification sound specification. The best results per model are highlighted in bold.}
  \label{tab:complete-results}
  \centering
  \begin{adjustbox}{max width=\textwidth}
  \newcolumntype{g}{>{\columncolor{blue!10}}c}
  \newcolumntype{b}{>{\columncolor{blue!10}}c}
  \begin{tabular}{l c bg g gg g gg g gg g gg g gg }
    \toprule
    \rowcolor{white}
    && &&  \multicolumn{3}{c}{\bfseries Bug complete} &&  \multicolumn{3}{c}{\bfseries Verification complete} &&  \multicolumn{1}{c}{\bfseries Mutants}\\
    \cmidrule{5-7} \cmidrule{9-11} \cmidrule{13-13} \rowcolor{white}
    Model && Prompt && \% complete && $x/164$ && \% complete && $x/164$ && \% killed \\
    \midrule
    \rowcolor{white}
    \multirow{ 2}{*}{GPT-5 (Chat)} && \NLPost && 7.3 && 15 && 4.8 && 11 && 11.3 \\
    && \NLC && \textbf{39.5} && \textbf{83}  && \textbf{29.9} && \textbf{68} && \textbf{73.0} \\ 
    \midrule
   \rowcolor{white}
    \multirow{ 2}{*}{GPT-4o} && \NLPost && 4.8 && 15 && 3.8 && 11 && 12.8 \\ 
    && \NLC && \textbf{31.4} && \textbf{92} && \textbf{25.2} && \textbf{69} && \textbf{73.7}\\
    \midrule
    \rowcolor{white}
    \multirow{ 2}{*}{CodeQwen 2.5} && \NLPost && 4.6 && 16 && 2.8 && 11 && 11.4\\
    && \NLC &&  \textbf{18.4} &&  \textbf{80} &&  \textbf{12.9} &&  \textbf{58} &&  \textbf{69.1}\\
    \bottomrule
  \end{tabular}
  \end{adjustbox}
    \vspace{-1em}
\end{table*}

 \inlineheadingbf{Results} \Cref{tab:complete-results} summarizes our experimental results. Overall, we find that for {\textsc{HumanEval}$^+$} mutants: 
 
 \inlineheadingit{LLMs can generate bug discriminating functional contracts from natural language} We find that LLMs prompted with \NLC can generate at least one verification sound functional contract that is able to kill all code mutants for 80 to 92 {\textsc{HumanEval}$^+$} tasks (48.8\% to 56.1\% of all tasks). Surprisingly, GPT-4o performs better in generating bug complete specifications for a diverse set of tasks, producing at least one verification sound bug complete functional contract for 92 tasks. However, our results also indicate that GPT-5 (Chat) is more reliable in generating verification sound bug complete specifications: 39.5\% of the GPT-5 (Chat) generated functional contracts are verification sound {\em and} bug complete. While the open-source model CodeQwen 2.5 achieves a lower bug completeness score, 18.4\% of the generated specifications are still verification sound and bug complete. When prompted with \NLPost, LLMs perform significantly worse producing a verification sound bug complete postcondition in 4.6\% to 7.3\% of all cases. This is expected as LLMs prompted with \NLPost are in general less effective in producing verification sound formal specifications (as we have seen in \textbf{RQ1}).
 
\inlineheadingit{Bug completeness overestimates verification completeness} Recall that completeness evaluates the effectiveness of specifications to catch code mutants via specification violations. Bug completeness evaluates completeness by testing the specification with a set of bug-triggering inputs. Verification completeness, in contrast, uses a software verifier to generate bug triggering inputs. Because of this difference, we find that bug completeness overestimates verification completeness in practice: LLMs produce less verification sound specifications that are useful for the software verifier to detect all code mutants. Still, LLMs prompted with \NLC are more effective in generating verification sound and verification complete specifications than LLMs prompted with \NLPost. In fact, between 12.9\% and 29.9\% of the LLM generated functional contracts are verification sound and verification complete while only 2.8\% to 4.8\% of \NLPost postconditions are verification complete. In addition, LLMs prompted with \NLC can generate verification sound functional contracts that let the software verifier detect all code mutants for 58 to 69 {\textsc{HumanEval}$^+$} tasks (35.4\% to 42.1\% of all tasks). 

\inlineheadingit{Mutation score}  We also evaluate the percentage of mutants killed by the verifier when provided with a verification sound specification. Our results are also shown in  \Cref{tab:complete-results} (right hand side). We find that the software verifier given a verification sound LLM-generated functional contract can discriminate over three-quarters of the buggy code mutants. 

\begin{figure}[t]
    \centering
    \begin{subfigure}{0.5\textwidth}
       \centering
         \includegraphics[width=\textwidth]{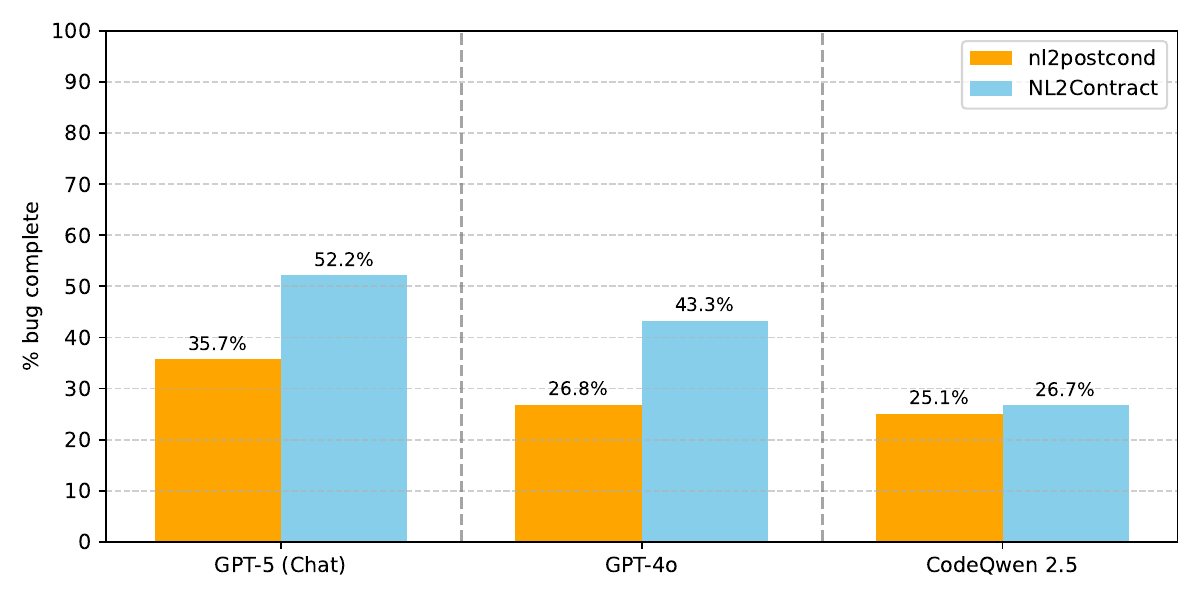}
  	\caption{Postcondition bug completeness}
  	\label{fig:post-bug-complete}
    \end{subfigure}\hfill
    \begin{subfigure}{0.5\textwidth}
        \centering
         \includegraphics[width=\textwidth]{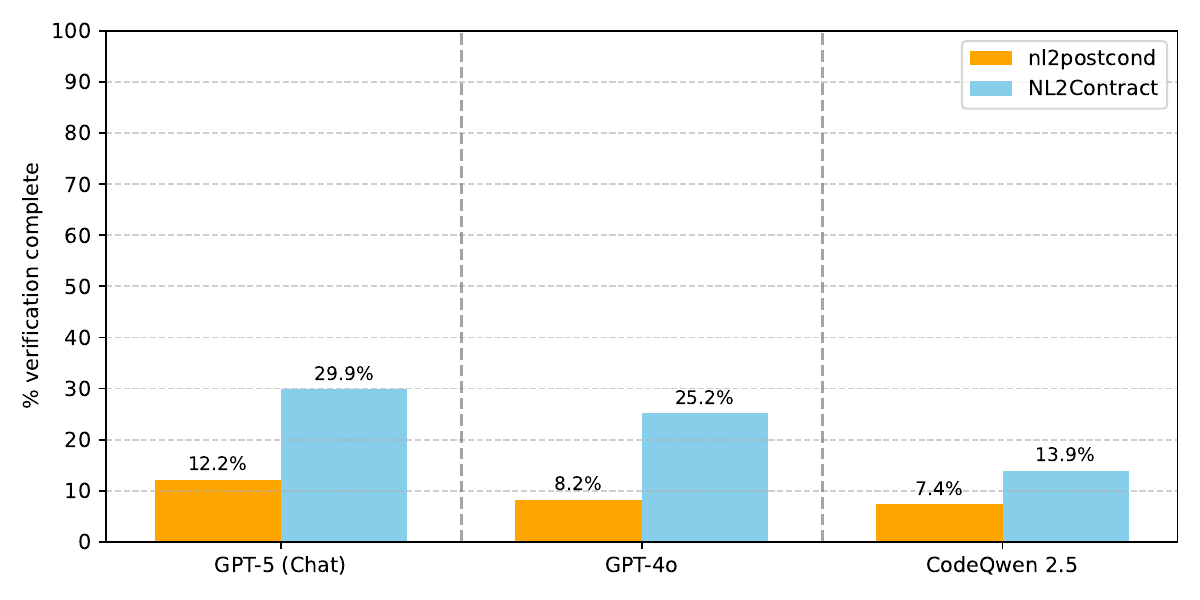}
  	\caption{Postcondition verification completeness}
  	\label{fig:post-verification-complete}
    \end{subfigure}
    \vspace{-1.5em}
    \caption{Comparison of LLMs under different prompts with respect to postcondition completeness. }
    \label{fig:post-complete}
 
  \vspace{-1em}
\end{figure}

\inlineheadingit{Postcondition completeness} For a fair comparison between \NLPost and \NLC, we evaluate the completeness of the generated postconditions. For this, we again add the reference precondition to the generated contracts as described before. We thus consider all contracts that are verification sound for {\em valid} inputs, i.e. those that are verification sound under the reference precondition. Our results for postcondition completeness are shown in \Cref{fig:post-complete}. In \Cref{fig:post-bug-complete}, we compare the bug completeness of the postconditions generated by LLMs prompted with \NLPost and \NLC. We find that LLMs prompted to generate functional contracts are more effective than the same LLMs prompted to generate postconditions (with up to 18.5\% more specifications being bug complete). This also carries over to verification completeness (see \Cref{fig:post-verification-complete}). Here, 13.9\% to 29.9\% of the postconditions generated by LLMs prompted for \NLC are verification complete while only 7.4\% to 12.2\% of postconditions generated by LLMs prompted for \NLPost are verification complete.

Based on our results, we conclude for \textbf{RQ2}:
\llbox{On the {\textsc{HumanEval}$^+$} benchmark, LLMs are more effective in generating functional contracts that can discriminate between buggy and correct behavior. In addition, when provided with \NLC contracts, the software verifier can {\em automatically} detect up to 73\% of all mutants, allowing an automatic verification of up to 69 out of 164 tasks.} 

\subsection{RQ3: Finding Real Bugs with Automatic Verifiers}\label{sec:rq3}
For \textbf{RQ3}, we evaluate whether LLM generated functional contracts can be used to identify real software bugs with the help of automatic verification techniques. We are not only considering formal verification tools, but we also evaluate the potential of the generated contracts for automatic testing. 

\inlineheadingbf{Python-by-Contract} To evaluate on real world bugs, we employ the {\em Python-by-Contract} dataset~\cite{zhang2022python}. The dataset collects Python solutions for 55 introductory tasks from Advent of Code 2020 and the introductory programming course at ETH Zurich in Fall 2019. Each solution is manually annotated with reference functional contracts. In addition, the authors provide 59 {\em buggy} solutions that represent mistakes made during the development or specification process. We find that 31 bugs are due to implementation mistakes in functions and methods from which 19 can be confirmed by CrossHair with the reference specification. We choose Python-by-Contract because (1) it provides {\em realistic} buggy solutions together with a reference implementation, (2) the tasks are often non-trivial implementing between 1 to 10 functions per problem, and most importantly (3) they provide a reference specification which we use to derive the ground truth preconditions from.  To construct the dataset, we pair each buggy solution with its reference implementation and a reference precondition computed from the reference specification. In the process, we remove the original specifications and all comments hinting at the bug location.

\begin{table*}
  \caption{Comparison of different LLMs and prompts on the Python-by-Contract dataset. \%detecting is the percentage of specifications that enable the verifier (or tester) to detect the bug by reporting a bug triggering specification violation. \#bugs is the number of bugs (out of 31) where at least one bug triggering input is found by the verifier or tester.}
  \label{tab:pbc-results}
  \centering
  \begin{adjustbox}{max width=\textwidth}
  \newcolumntype{g}{>{\columncolor{blue!10}}c}
  \newcolumntype{b}{>{\columncolor{blue!10}}c}
  \begin{tabular}{l c bg g gg g gg g gg g gg g gg }
    \toprule
    \rowcolor{white}
    && && &&  \multicolumn{3}{c}{\bfseries CrossHair} &&  \multicolumn{3}{c}{\bfseries Pynguin} \\
    \cmidrule{7-9} \cmidrule{11-13} \rowcolor{white}
    Model && Prompt && $sound@1$  && \% detecting  && \# bugs && \% detecting  && \# bugs  \\
    \midrule
    \rowcolor{white}
    \multirow{ 2}{*}{GPT-5 (Chat)} && \NLPost && 20.0 && 4.2 && 3 && \textbf{16.1} && 6 \\
    && \NLC && \textbf{71.3} && \textbf{39.4}  && \textbf{14} && 15.8 && \textbf{8} \\
    \midrule
   \rowcolor{white}
    \multirow{ 2}{*}{GPT-4o} && \NLPost && 19.7 && 5.8  && 5  && \textbf{15.8} && \textbf{9} \\
    && \NLC && \textbf{67.7} && \textbf{34.8} && \textbf{14} && 14.8 && 7 \\
    \midrule
    \rowcolor{white}
    \multirow{ 2}{*}{CodeQwen 2.5} && \NLPost && 21.0 &&  6.8 && 4 && \textbf{16.1} && \textbf{7}\\
    && \NLC &&  \textbf{64.8} &&  \textbf{35.2} &&  \textbf{14} && 15.2 && \textbf{7}\\
    \bottomrule
  \end{tabular}
  \end{adjustbox}
    \vspace{-1em}
\end{table*}

\inlineheadingit{Experimental setup} We compare specifications generated by LLMs prompted with \NLC and \NLPost. To construct the prompts, we follow the process described in \Cref{sec:contract}. In particular, we remove the body of all functions and methods implemented in the buggy solution and prompt the LLMs based on the provided natural language hints (that are already available in code). LLMs are prompted to generate specifications for functions that are modified in the solution process. We use the reference implementation and precondition to compute the $sound@1$ score. To find real bugs automatically, we employ different {\em automatic} tools with the generated specifications. In the process, we say that a bug is detected if the tool given a generated specification reports a {\em bug triggering} violation $\sigma_{\text{\faTimes}}$ on the buggy solution. A violation is bug triggering if the buggy implementation behaves differently than the reference implementation for the given input, i.e.  $\Pre_{gt}(\sigma_\text{\faTimes}) \land f_P(\sigma_\text{\faTimes}) \neq f_{gt}(\sigma_\text{\faTimes})$. 

\inlineheadingit{Verifiers and Testers} We are not only considering CrossHair as our main verifier, but also evaluate Pynguin~\cite{lukasczyk2022pynguin} which is a popular automatic testing tool for Python. Pynguin implements a search-based coverage-guided testing strategy and is typically used
for generating regression tests. To utilize Pynguin, we exploit the properties of our functional contracts (see \Cref{sec:contract}): The contract defines a contract function that only raises assertion errors when the test input leads to a specification violation. Hence, when the program is instrumented with the generated contract function, we can detect specification violations by running the contract function on different test inputs. In this case, we employ Pynguin to generate random test inputs and report any test case that leads to an assertion error for the contract function. We run Pynguin with a 600s timeout.

\inlineheadingbf{Results} Our results are shown in \Cref{tab:pbc-results}. Overall, we find that for the Python-by-Contract bugs:

\inlineheadingit{LLMs can generate specifications that help to catch real bugs automatically} Supplied with \NLC generated specifications, CrossHair is able to automatically detect 14 out of 19 bugs that it is capable to find. In general, the LLMs prompted with \NLC produce in 34.8\% to 39.4\% of all cases a specification that enables the verifier to detect the bug. GPT-5 (Chat) is more effective than the other LLMs in generating sound specifications, which more often enables generation of a bug triggering inputs. CodeQwen 2.5 is surprisingly competitive with the other larger LLMs, allowing the detection of the same number of bugs with a model that can run locally.

\begin{figure}[t]
    \centering
    \begin{subfigure}{0.5\textwidth}
       \centering
         \includegraphics[width=\textwidth]{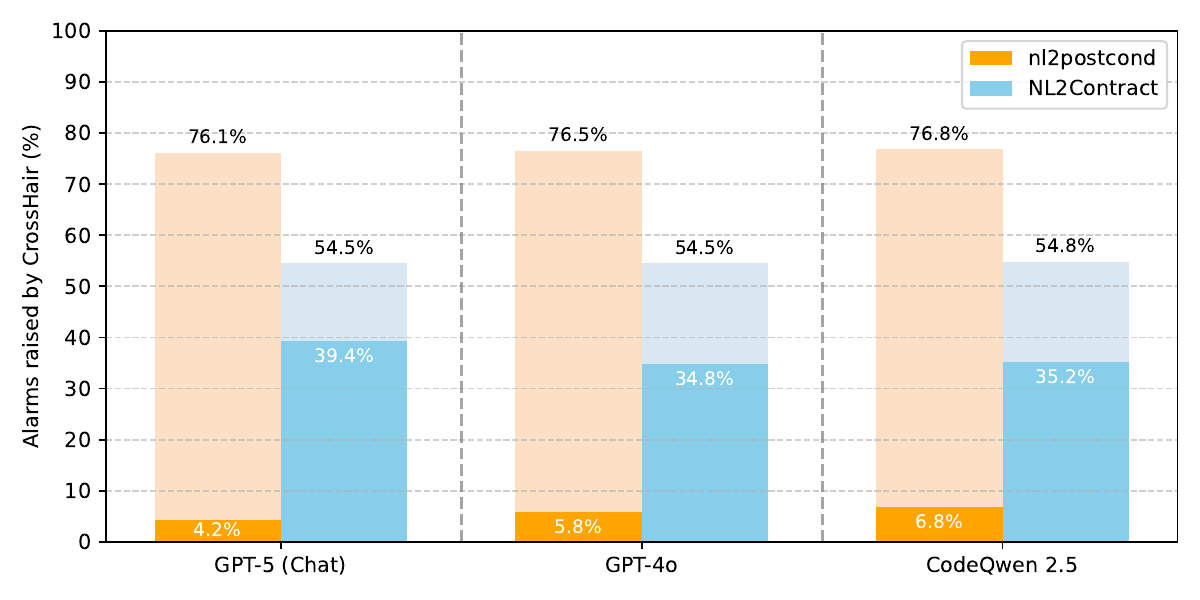}
  	\caption{Alarms raised by CrossHair. }
  	\label{fig:false-alarm}
    \end{subfigure}\hfill
    \begin{subfigure}{0.5\textwidth}
        \centering
         \includegraphics[width=\textwidth]{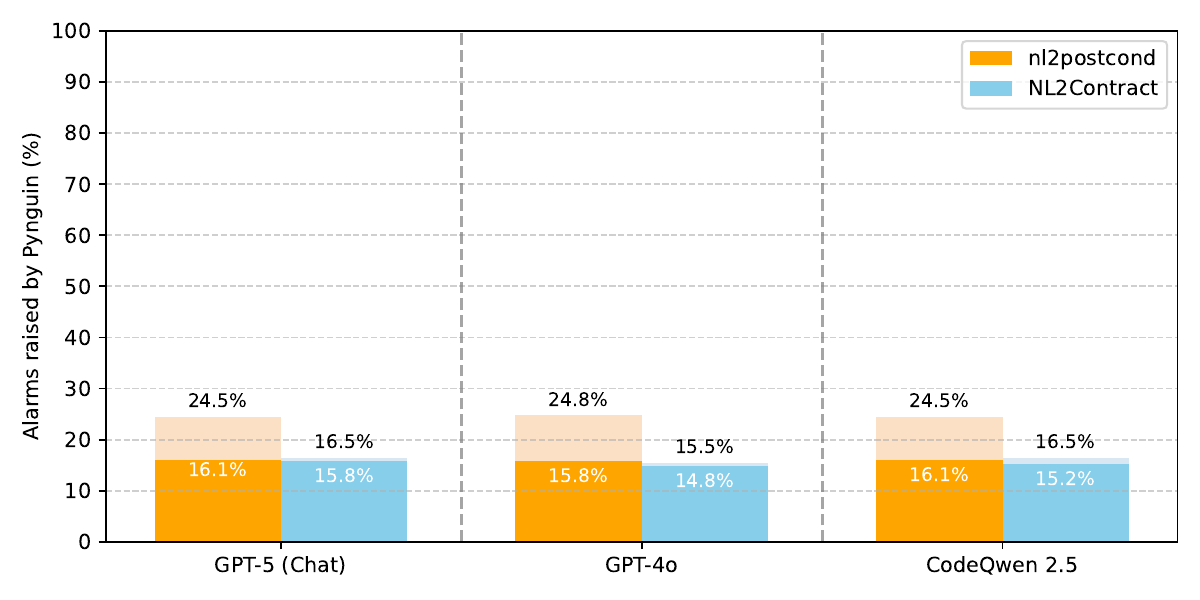}
  	\caption{Alarms raised by Pynguin. }
  	\label{fig:false-alarm-pynguin}
    \end{subfigure}
    \vspace{-1.5em}
    \caption{Analysis of the impact of LLM generated specification on the performance of automatic tools. The lighter color shows the percentage of alarms raised by the different tools. The darker color shows the percentage of alarms (specification violations) that identify a real bug.}
    \label{fig:bug-alarms}
 
  \vspace{-1em}
\end{figure}

\inlineheadingit{Functional contracts are necessary for avoiding false alarms} In comparison with \NLPost postconditions, LLMs with \NLC produce specifications that are significantly more useful in software verification: The software verifier supplied with \NLC based specifications finds 9 to 11 bugs more than when supplied with \NLPost postconditions. The key reason for this is the higher verification soundness of the \NLC specifications. Therefore, in practice, verifiers supplied with \NLC produce a significant lower number of false alarms, as shown in \Cref{fig:false-alarm}. Here, we report the percentage of specifications that lead to an alarm raised by CrossHair on the buggy solutions. We find that most alarms reported by CrossHair given the \NLPost specifications are {\em false alarms}, i.e. specification violations that are not bug triggering. While CrossHair also reports false alarms with \NLC specifications, most of the found specification violations uncover a real bug. This showcases the higher usability of \NLC specifications in the context of software verification.

\inlineheadingit{Contract functions for random testing} Recall that we express functional contracts as contract functions which can be evaluated by executing the code on different input. We utilize this property to evaluate whether the generated contract functions are also useful for automated random testing (right hand side of \Cref{tab:pbc-results}). By injecting the contract functions into the program, we find that Pynguin (without any modification) can effectively find up to 9 out of 31 bugs. LLMs prompted with \NLPost perform slightly better in the context as Pynguin spends a significant amount of its time in covering the \NLC preconditions. Still, when Pynguin finds a specification violation, we find that they are more often bug triggering for \NLC contracts, as shown in \Cref{fig:false-alarm-pynguin}. We see this as an indication that \NLC contracts can be promising in the context of automated testing, but we leave the exploration of specification formats tailored for testing and evaluations with property-based testing methods~\cite{DBLP:journals/jossw/MaciverH19} open for future work.

Finally, we conclude for \textbf{RQ3}:
\llbox{For Python-by-Contract bugs, the \NLC specifications are more {\em useful} for finding
bugs automatically with software verifiers than raw \NLPost postconditions. Software verifiers supplied with \NLC specifications can not only detect up to 14 out of 31 bugs automatically, but also produce significantly less false alarms in the process. }

\section{Discussion}

\begin{figure}
\centering
\adjustbox{max width=\textwidth}{
\tikzset{
  warningbox/.style={
    draw=red!80!black,
    fill=red!5,
    very thick,
    rounded corners,
    align=left,
    text width=0.55\linewidth,
    inner sep=6pt,
  },
}

\begin{tikzpicture}

%
%
%

\node[anchor=north west] (code) [fill=gray!8!white, rounded corners, inner sep=2pt, text width=7.5cm] at (0,0) {%
\begin{python}[escapechar=!, numbers=none, basicstyle=\ttfamily\tiny, xleftmargin=2pt]
def compute_angles(hour: int, minute: int, second: int) 
		-> Tuple[float, float, float] :
    """Compute the angles of the clock hands for a given time of the day."""
    !\color{black!40!white}angle\_second = second / 60 * 360!
    !\color{black!40!white}angle\_minute = (minute + second / 60) / 60 * 360!

    !\color{comment!40!white}\# ERROR!
    !\color{comment!40!white}\# I forgot to distinguish between <12 and >= 12 clock hours.!
    
    !\color{black!40!white}angle\_hour = (hour + minute / 60 + 
    					second / 3600) / 12 * 360!

    !\color{black!40!white}return angle\_hour, angle\_minute, angle\_second!

\end{python}
};

\node[right=0.5cm of code] (format) [fill=gray!8!white, rounded corners, inner sep=8pt]  {%
\adjustbox{max width=0.8\textwidth}{%
\begin{python}[escapechar=!, numbers=none, basicstyle=\ttfamily\tiny]
def compute_angles_contract(hour: int, minute: int, second: int) 
	-> Tuple[float, float, float]:
    try:
        # Preconditions
        assert 0 <= hour <= 23
        assert 0 <= minute <= 59
        assert 0 <= second <= 59
    except AssertionError as e:
        raise ValueError(f"Precondition failed: {e}") from e
    result = compute_angles(hour, minute, second)
    ...
    # Check angles are within valid bounds
    for angle in result: assert 0.0 <= angle < 360.0
    ...
    return result
\end{python}
}};
\node [below=0.2cm of format] (format-header) {(b) \texttt{Function Contract}};
\node[below right=-0.6cm and -4.6cm of format] (mutant1) [fill=red!8!white, rounded corners, inner sep=6pt] {%
\faTimes ~ Violation when \texttt{hour == 12}
};

\node [left=5cm of format-header] (code-header) {(a) \texttt{Buggy solution}};

\end{tikzpicture}
}
\vspace{-1.5em}
\caption{Example of buggy solution from the Python-by-Contract dataset and a simplified contract generated by GPT-5 (Chat) with \NLC. The grayed out part is {\em not} provided to the LLM. Given the specification, CrossHair correctly identifies the bug when calling \texttt{compute\_angles(12, 0, 0)}.}\label{fig:bug-ex}
\vspace{-1em}
\end{figure}

To gain further insights into how \NLC function contracts enable the detection of real bugs with automatic tools, we conduct a qualitative evaluation of specifications generated by LLMs that led to successful bug detection and that allow false alarms. We discuss two cases in the following that are representative for the strengths of \NLC and LLM-based specification inference in general.

\inlineheadingbf{Preconditions are important for real bug detection} While we already showcased an example where preconditions matter for bug detection in our introduction, we now consider a real bug from the Python-by-Contract dataset in \Cref{fig:bug-ex}. The goal of the benchmark task is to implement a conversion from a 24h clock into angles of clock hands. \Cref{fig:bug-ex}(a) shows a buggy solution where the implementer forgot to handle the time after 12 clock hours and \Cref{fig:bug-ex}(b) provides the contract generated by GPT-5 (Chat) which we simplified for demonstration. During specification inference, the LLM is only provided with the function signature and docstring, while the implementation is hidden from the LLM. Given the specification, CrossHair correctly reports a bug triggering specification violation for the input \texttt{compute\_angles(12, 0, 0)}.  We find that this examplifies a case where generating preconditions is of particular importance: The precondition ensures that all inputs follow the format of 24h clock. GPT-5 (Chat) prompted with \NLPost does not encode this information in the generated postcondition: 
\begin{center}
\vspace{-0.5em}
\begin{python}[numbers=none]
                   assert all(0.0 <= angle < 360.0 for angle in return_value)
\end{python}
\vspace{-0.5em}
\end{center}
Although the postcondition is bug discriminating for valid inputs, CrossHair reports a postcondition violation for the input \texttt{compute\_angles(-1, -1, -1)}. However, the behavior of \texttt{compute\_angles(-1, -1, -1)} is undefined which results into a false alarm, i.e. the buggy solution and reference implementation behave identical for the given input. In practice, such bug reports by automatic tools are unhelpful and might lead the developer away from detecting the real bug. Therefore, to really support the developer in bug detection, we need tools that only report bugs for valid inputs. As determining the validity of inputs is non-trivial, the example demonstrates the importance of preconditions for real bug detection with automatic verification tools and the potential of \NLC to generate sound preconditions from natural language hints. 

\inlineheadingbf{LLM inferred specifications can help to uncover inconsistency bugs automatically} During our investigation, we found an inconsistency bug which we reported to the Python-by-Contract dataset developers and which is shown in \Cref{fig:inconsistency}. In this example, the developer implements an approximation of the \texttt{sqrt} function which should be precise up to \texttt{eps}.  Both \NLC and \NLPost correctly capture this property with the postcondition \mbox{\pythoninline{abs(result * result - c) < eps}}. CrossHair however reports a specification violation for the reference implementation because the reference implementation only ensures that \pythoninline{abs(result * result - c) <= eps} holds, i.e. it computes  \pythoninline{approximate_sqrt(2, 0.25) = 1.5} which is too imprecise. This showcases the potential of specifications inferred from natural language descriptions. When confronted with this inconsistency bug, the developer can decide whether it represents a simple mistake in the documentation or a serious implementation flaw. This can help the developer to make the natural language description more consistent with the implementation or fix the bug. Both tasks are important for designing effective and reliable software systems~\cite{tan2007icomment}. Overall, we see \NLC as an important step towards verifying software with respect to natural language specifications already available in code which cannot only help us to uncover real bugs, but also inconsistencies with the documentation.


\begin{figure}
\centering
\adjustbox{max width=\textwidth}{
\tikzset{
  warningbox/.style={
    draw=red!80!black,
    fill=red!5,
    very thick,
    rounded corners,
    align=left,
    text width=0.55\linewidth,
    inner sep=6pt,
  },
}

\begin{tikzpicture}

%
%
%

\node[anchor=north west] (code) [fill=gray!8!white, rounded corners, inner sep=2pt, text width=7.5cm] at (0,0) {%
\begin{python}[escapechar=!, numbers=none, basicstyle=\ttfamily\tiny, xleftmargin=2pt]
"""
Approximate the square root of non-zero positive integer ``c``...
The result should be precise up to ``eps``: ``abs(t*t -c) < eps``.
"""
# Reference postcondition: abs(result * result - c) !\colorbox{red!80}{\color{white}<=}! eps
def approximate_sqrt(c: int, eps: float) -> float:
  """Approximate the square-root of c up to the precision eps."""
\end{python}
};
\node [below=0cm of code] (code-header) {(a) Reference docstring and function stub};

\node[above right=-0.75cm and 0.5cm of code] (format) [fill=gray!8!white, rounded corners, inner sep=4pt, text width=6.3cm]  {%
\adjustbox{max width=\textwidth}{%
\begin{python}[escapechar=!, numbers=none]
   assert abs(result * result - c) < eps 
\end{python}
}};
\node [below right=0cm and -6.6cm of format, text width=7cm] (format-header) {(b) \texttt{Relevant postcondition generated by \NLC and \NLPost}};
\node[below right=0cm and -7.25cm of format-header] (mutant1) [fill=red!8!white, rounded corners, inner sep=6pt, text width=6.1cm] {%
\small\faTimes ~ Violation when \texttt{c == 2} and \texttt{eps == 0.25}
};
\node [below=0cm of mutant1, text width=6cm] (format-header) {(c) \texttt{Violation reported by CrossHair}};

\end{tikzpicture}
}
\vspace{-1.5em}
\caption{An inconsistency bug in the Python-by-Contract dataset. }\label{fig:inconsistency}
\vspace{-1em}
\end{figure}

\section{Threats to Validity}
We have conducted our evaluation on two popular benchmarks: {\textsc{HumanEval}$^+$} and Python-by-Contract. Although used frequently in the evaluation of LLMs (especially {\textsc{HumanEval}$^+$}), the benchmarks focus on simple Python programs which often do not have dependencies beyond the scope of a single file. Therefore, our evaluation results might not generalize to other more complex (and potentially undocumented) real world Python code. Another potential threat is the risk of data leakage.
Because the Python-by-Contract dataset and {\textsc{HumanEval}$^+$} are public benchmarks, the underlying LLM might have seen and memorized the benchmark tasks. Our evaluation partially mitigates this risk by generating contracts in custom Python-based format. Python-by-Contract uses a more specific annotation language and we are not aware of a dataset with complete specification-annotated HumanEval tasks. For Python-by-Contract, there is still the risk that we measure the transfer of memorized specifications to our custom format.  Another potential limitation is the underlying software verifier. The software verification community has in the past mostly focused on statically typed compiled languages such as C or Java. Therefore, software verification tools for more dynamic languages such as Python are currently not as mature as their static counterpart. Our evaluation results still show that using an existing software verifier can help for finding real bugs in Python programs. We still expect that usability of \NLC contracts will increase with future, more mature verification tools for Python.
\section{Related Work}
In this work, we revisit the problem of specification inference from natural language descriptions, already available in code. In the process, we propose the task \NLC which evaluates the effectiveness of LLMs to generate sound and bug discriminating specifications useful for software verification. In the following, we discuss the most closely related approaches. 

\inlineheadingbf{Specification Generation} Formal specifications play a critical role in many areas of software engineering, including code generation~\cite{DBLP:conf/hase/WhalenH99, DBLP:conf/fmcad/AlurBJMRSSSTU13, DBLP:conf/cav/LiPP24}, software testing~\cite{DBLP:conf/psse/Gaudel07, DBLP:journals/jossw/MaciverH19, DBLP:journals/corr/MeerKH14}, and formal verification~\cite{beyer2025improvements, DBLP:conf/popl/BallR02, DBLP:conf/ershov/KhoroshilovMPZ09,  DBLP:conf/cav/BeyerK11, DBLP:conf/tacas/HeizmannCDGHLNM18, DBLP:conf/kbse/PavlinovicLS16}. They are used to describe the intended behavior of a system, including functional relationships of input and outputs, as well as invariants over the internal state. 
However, writing or obtaining formal specifications for a system under test is generally considered a hard task~\cite{DBLP:conf/fmics/DavisCCFHHHMW13}. Therefore, there exists a large body of research automating the generation of formal specifications~\cite{DBLP:conf/popl/0001NMR16, DBLP:conf/iclr/RyanWYGJ20, DBLP:conf/nips/SiDRNS18, DBLP:conf/fase/JanssenRW24, DBLP:conf/sigsoft/DinellaLN24, DBLP:conf/pldi/AstorgaMSWX19, DBLP:conf/sefm/GhosalJR23, DBLP:conf/kbse/MolinadA23, DBLP:conf/icse/WeiFKM11, DBLP:series/lncs/AlshnakatGLR20}. Many existing works focus on loop invariant generation~\cite{DBLP:conf/popl/0001NMR16, DBLP:conf/iclr/RyanWYGJ20, DBLP:conf/nips/SiDRNS18, DBLP:conf/fase/JanssenRW24}, while others have attempted the generation of preconditions~\cite{DBLP:conf/sigsoft/DinellaLN24, DBLP:conf/pldi/AstorgaMSWX19, DBLP:conf/sefm/GhosalJR23}, postconditions~\cite{DBLP:conf/kbse/MolinadA23, DBLP:conf/icse/WeiFKM11, DBLP:series/lncs/AlshnakatGLR20}, and assertion-based invariants~\cite{DBLP:conf/sigsoft/TerragniJTP20}. While useful for building regression oracles~\cite{DBLP:conf/icst/YuQAR13}, most of these techniques focus on capturing the existing implemented behavior, instead of the intended behavior, limiting their application for bug detection. Our goal in contrast is to generate formal specifications of the intended program's behavior from natural language descriptions, which are useful for bug detection via verification. Several approaches have been proposed that, similar to our work, aim to generate specifications from natural language~\cite{blasi2018translating, tan2007icomment, tan2012tcomment, DBLP:conf/sigsoft/ZhaiSPZLFM0020, DBLP:journals/jss/BlasiGEPC21, DBLP:journals/ase/ZhongZXM11, DBLP:conf/issta/KimCSOPTC23, DBLP:conf/icse/PanditaXZXOP12}. Many of these techniques rely on pattern-matching and hand-crafted rules. For example, Jdoctor~\cite{blasi2018translating}, icomment~\cite{tan2007icomment}, and @tComment~\cite{tan2012tcomment} use pattern-matching techniques to extract specifications from Javadoc comments. This often restricts their application to semi-structured natural language formats, making their adoption for general specification inference from natural language hints in code challenging. In contrast, \NLC utilizes the code understanding capabilities of LLMs to infer functional contracts from the few natural language hints that are already available in code.

\inlineheadingbf{LLMs for Specification Generation} Several approaches have been proposed to synthesize program specifications using LLMs~\cite{ma2024specgen, wen2024enchanting, endres2024can, dinella2022toga, hossain2024togll, tufano2020unit, DBLP:conf/ast/TufanoDSS22, DBLP:conf/ast/PrimbsFF25}. Most of these approaches are predominantly used for testing. 
AthenaTest~\cite{tufano2020unit} uses machine learning to generate both the input and the oracle of a unit test. TOGA~\cite{dinella2022toga}, TOGLL~\cite{DBLP:conf/icse/HossainD25}, and Doc2OracLL~\cite{DBLP:journals/pacmse/HossainTD25} generate exception and assertion based test oracles. TiCoder~\cite{lahiri2022interactive} leverages LLMs to generate test inputs and expected outputs. While useful for testing, the generated test oracles are often limited to describing the expected behavior for a specific test input (test prefix). Our goal are specifications of the expected behavior that are valid for all inputs. There also exist more recent approaches that utilize LLMs for generating formal specifications such as AutoSpec~\cite{wen2024enchanting} and SpecGen~\cite{ma2024specgen}. They use LLMs in combination with automatic software verifiers to infer and refine specifications until the specification is consistent with the current implementation. As a consequence, the generated specifications capture the implemented behavior of the existing (buggy) code precisely, making bug detection impossible. In contrast, we propose \NLC as a way to measure the ability of LLMs to infer the desired and expected behavior from natural language. Our experiments demonstrates that functional contracts derived from natural language hints allows to find novel bugs, which are inconsistencies between the documentation and the actual implementation, with the help of automatic verification tools. Most closely related to our work is \NLPost~\cite{endres2024can} which evaluates the ability of LLMs to generate postconditions describing the intended code behavior from natural language elements. \NLC extends this task to the generation of formal contract. Our experiments show that LLMs prompted with \NLC can generate sound functional contracts which are more useful than raw postconditions for the detection of real bugs with an automatic software verifier.

\section{Conclusion}
In this work, we introduce \NLC as the task to infer functional contracts from natural language descriptions via LLMs. We evaluate
the capabilities of LLMs to generate sound functional contracts -- consisting of both pre- and postconditions -- useful for
finding software bugs with automatic verification tools. We formally define key quality indicators of the specifications such as verification soundness
and verification completeness that we use for evaluating the quality of the generated specifications. Our evaluation shows that LLMs
can translate natural language descriptions into non-trivial contracts that accurately capture the developer's intent. Our study further reveals
that LLMs can generate meaningful preconditions which allows an automatic software verifier to find 14 real world bugs in the Python-by-Contract
benchmark, while emitting significantly fewer false alarms. Our research shows the potential of LLMs to generate formal specifications from natural language, which ultimately
can enable software validation and bug detection with automatic verification tools.

\section{Data Availability}
We plan to make all implementations and data will publicly available and archived at Zenodo. 

\bibliographystyle{ACM-Reference-Format}
\setcitestyle{numbers,sort&compress}
\bibliography{literature}

\end{document}